\documentclass[twocolumn]{aastex631}

\shorttitle{Mutual Validation of Tilt Angles Datasets}
\shortauthors{Qin et al.}

\graphicspath{{./}{figures/}}

\begin{document}

\title{Mutual Validation of Datasets for Analyzing Tilt Angles in Solar Active Regions}
\correspondingauthor{Jie Jiang}
\email{jiejiang@buaa.edu.cn}

\author{Lang Qin}
\affiliation{School of Space and Earth Sciences, Beihang University, Beijing, People’s Republic of China}
\affiliation{Key Laboratory of Space Environment Monitoring and Information Processing of MIIT, Beijing, People’s Republic of China}

\author{Jie Jiang}
\affiliation{School of Space and Earth Sciences, Beihang University, Beijing, People’s Republic of China}
\affiliation{Key Laboratory of Space Environment Monitoring and Information Processing of MIIT, Beijing, People’s Republic of China}

\author{Ruihui Wang}
\affiliation{School of Space and Earth Sciences, Beihang University, Beijing, People’s Republic of China}
\affiliation{Key Laboratory of Space Environment Monitoring and Information Processing of MIIT, Beijing, People’s Republic of China}

\begin{abstract} 
The tilt angle of solar active regions (AR) is crucial for the Babcock-Leighton type dynamo models in the buildup of polar field. However, divergent results regarding properties of tilt angles were reported due to their wide scatter, caused by intrinsic solar mechanisms and measurement errors. Here, we mutually validate the magnetogram-based AR tilt angle dataset from Wang, Jiang, \& Luo with the Debrecen Photoheliographic Data by identifying common data points where both datasets provide comparable tilt angles for the same AR/sunspot. The mutually validated datasets effectively reduce measurement errors, enabling a more accurate analysis of the intrinsic properties of tilt angles. Our mutually validated datasets reveal that the difference between white-light-based and magnetogram-based tilt angles has no significant difference. Also, the datasets show that an upward revision of average tilt angle  ($\bar\alpha$) and a downward revision of the tilt scatter ($\sigma_\alpha$) compared to previous results are necessary, with typical values of about 7$^\circ$ and 16$^\circ$, respectively. The $\sigma_\alpha$ values demonstrate a strong correlation with AR flux and sunspot area, with the dependency functions re-evaluated using mutually validated datasets. Furthermore, both $\bar\alpha$ and the tilt coefficient for the weak cycle 24 are larger than those for cycle 23. This supports the tilt quenching mechanism, which posits an anti-correlation between the cycle-averaged tilt angle and cycle amplitude. Additionally, tilt angle from the mutually validated dataset has a weak non-monotonic relationship with magnetic flux and does not depend on the maximum magnetic field strength of ARs. 
\end{abstract}


\section{Introduction} \label{sec:intro}
The line connecting the two polarities of a sunspot group or active region (AR) slightly tilts with the solar equator \citep{Hale1919}. This AR tilt angle contributes to the net dipole field and is a fundamental aspect of the Babcock-Leighton (BL) mechanism \citep{Babcock1961, Leighton1969}, which describes the generation of poloidal magnetic fields as a part of the solar dynamo loop \citep{Jiang2013, Charbonneau2020}. Investigating the properties of these tilt angles is crucial for understanding the solar cycle, as they provide insights into the dynamo processes within the BL-type dynamo framework. Moreover, the tilt angle, as an intrinsic characteristic of ARs, offers valuable information about the flux emergence process \citep{Fan2021, Weber2023}, which cannot be directly observed and remains a significant open question in solar physics. 

The average value of tilt angle, $\bar\alpha$, indicates the efficiency of poloidal field generation from the toroidal field \citep{Wang1991}. However, previous studies reported a wide range of values for $\bar\alpha$, with an opinion that tilt angles measured using white-light observations were systematically lower than those measured using magnetograms \citep{Howard1996a, Wang2015}. \cite{Howard1996b} presented an example where the value of $\bar\alpha$ based on Mount Wilson Observatory (MWO) white-light data over the period 1917-1985 was 4$^{\circ}$.3, whereas the corresponding magnetogram-based data during 1967-1995 was 6$^{\circ}$.3. \cite{Dasi2010} reported the values of $\bar\alpha$ for MWO and Kodaikanal (KK) white-light measurements as 4$^{\circ}$.25 and 4$^{\circ}$.5, which is consistent with \cite{Howard1996b}. However, \cite{Jiao2021} showed that with an angular separation constraint to remove unipolar spots, the value of $\bar\alpha$ for KK sunspot tilts increases from $4^{\circ}.67$ to $7^{\circ}.01$. Using drawings of sunspot groups, \cite{Brunner1930} obtained a value of $\bar\alpha=6^{\circ}.5$. Based on the Pulkovo database of sunspot group measurements, \cite{Ivanov2012} reported $\bar\alpha=6^{\circ}.1$.  These white-light measurements are actually in line with the magnetic measurements given by \cite{Howard1996b}. Meanwhile, other magnetic measurements also reported a wide range of $\bar\alpha$. From the magnetograms recorded by the National Solar Observatory at Kitt Peak (NSO/KP) during cycle 21, \cite{Wang1989} obtained a large value of $\bar\alpha=9^{\circ}.0$. However, based on the MDI/SOHO and the HMI/SDO magnetograms, \cite{Li2018} obtained a small value of $\bar\alpha=4^{\circ}.58$. As the axial dipole field strength of a bipolar magnetic region is approximately proportional to its tilt angle \citep{Yeates2023}, the change of tilt angle from $4^\circ.3$ to $9^\circ.0$ roughly results in an axial dipole strength 2.1 times of its original value. Therefore, accurately measuring the tilt angle is critical.

Besides the systematic average value, there is a significant scatter in the tilt angles, quantified by the standard deviation, $\sigma_\alpha$. The amplitude of $\sigma_\alpha$ provides important insights: it constrains the effect of convective buffeting on flux emergence \citep{Weber2013, Schunker2019} and plays a crucial role in determining whether the variability of the solar cycle is driven by stochastic or chaotic mechanisms \citep{Olemskoy2013, Cameron2017, Jiang2020, Karak2023, WangZF2025}. According to \cite{Ivanov2012} (Table 1) and \cite{Jiao2021} (Table 4), $\sigma_\alpha$ of MWO and KK white-light measurements is close to 30$^{\circ}$, over six times the average value $\bar\alpha$ ($r = \sigma_\alpha / \bar\alpha > 6$). This ratio $r$ can be decreased to about 4 with the angular separation constraint \citep{Jiao2021}. Based on the Pulkovo database of sunspot group measurements, \cite{Ivanov2012} reported $\sigma_\alpha = 14^{\circ}$, with $r \approx 2.3$. While the $r$-value remains controversial, there is consensus on the decrease of $\sigma_\alpha$ with an increase in sunspot area \citep{Howard1991, Fisher1995, Jiang2014} or AR flux \citep{Wang1989, Stenflo2012, Jha2020}. Accurate $r$-value and the function describing the relation between $\sigma_\alpha$ and the sunspot area/flux are prerequisites for understanding the stochastic mechanism in solar cycle variability within the framework of the BL-type dynamo.

To account for solar cycle variability, tilt quenching is recognized as an efficient nonlinear mechanism \citep{Jiang2020, Talafha2022}. Observationally, two types of tilt quenching have been reported. The first type (TQ1) is an anti-correlation of tilt angle and cycle amplitude, firstly reported by \cite{Dasi2010}. Subsequent studies using different datasets once questioned its existence \citep{Ivanov2012, Tlatova2018, Isik2018}. The second type (TQ2) is a non-monotonic relationship between tilt angle and magnetic properties of ARs. \cite{Jha2020, Sreedevi2024} reported that tilt angle initially increases with increasing $B_{max}$ of ARs but decreases beyond a certain value. Nonetheless, there is a long debate about the existence and the specific magnetic properties involved in this relationship. For example, \cite{Tian2003} observed a non-monotonic relationship where the tilt angle depends on magnetic flux instead of $B_{max}$. \cite{Wang1989, Li2018} suggested that tilt angle is weakly anti-correlated with magnetic flux. In contrast, \cite{Stenflo2012} reported no correlation between magnetic flux and tilt angle. 

The divergent results regarding the properties of tilt angle \citep[for more details,  see][]{vanDriel-Gesztelyi2015} stem from its significant scatter, which originates from both intrinsic solar mechanisms and measurement errors. Addressing this scatter is crucial to accurately determine each tilt angle property. \cite{Jiao2021} made the first attempts to reduce the effects of scatter on the uncertainty of tilt properties. They evaluated the uncertainties of methods used in earlier studies and proposed that performing a linear fit to the tilt-latitude relation of sunspot groups, with an angular separation constraint and without binning the data, can minimize the effect of tilt scatter. They verified the existence of TQ1. In this paper, we aim to introduce a method to reduce measurement errors and enhance the reliability of tilt data by mutually validating a white-light-based dataset with a magnetogram-based dataset. This approach enables a more accurate analysis of the true properties of tilt angles based on the mutually validated data.

This paper is organized as follows. In Section \ref{sec:data}, we introduce the original datasets and describe the mutual validation method used. Section \ref{sec:result} investigates properties of tilt angles, including average values, scatter and its dependence on sunspot area/flux, and tilt quenching, based on the original and the mutually validated datasets. Finally, we discuss and conclude our results in Sections \ref{sec:discussion} and \ref{sec:conclusion}, respectively.

\section{Datasets and methods} \label{sec:data}
\subsection{Original datasets} \label{subsec:oridata}
\citet{Wang2023, Wang2024} provide a homogeneous database (hereafter WJL dataset) of solar active regions based on SOHO/MDI and SDO/HMI synoptic magnetograms. It provides heliographic positions, areas, and fluxes of both polarities of ARs. The magnetic field strength and flux of HMI magnetograms are calibrated by multiplying a factor of 1.36. The WJL dataset does not provide tilt angles. Therefore, we calculate the tilt angle $\alpha$ of ARs using the equation:
\begin{equation}
	\tan\alpha = \frac{\Delta\lambda}{\Delta\phi\cos\lambda},
	\label{eq:tilt}
\end{equation}
where $\lambda$, $\Delta\lambda$, and $\Delta\phi$ are the heliographic latitude, latitudinal and longitudinal angular separations, respectively. The centroids of the positive and negative polarities are calculated using flux-weighting. This method is the traditional and widely accepted approach for determining the tilt angle \citep{Howard1991}. We use data from 2552 ARs spanning solar cycles 23 and 24, with 1453 ARs from cycle 23 and 1099 ARs from cycle 24.

Debrecen Photoheliographic Data (DPD) sunspot catalogue\footnote{http://fenyi.solarobs.epss.hun-ren.hu/test/tiltangle/dpd/} provides heliographic positions, areas, central meridian distance (CMD) and tilt angles of whole sunspot groups and pores, from January 1974 to January 2018 based on white-light observations \citep{Baranyi2015, Baranyi2016, Gyori2017}. The data post-2015 was personally provided by Tünde Baranyi in 2018 \citep{Jiang2019}. The tilt angles were calculated using the same formulation as Equation (\ref{eq:tilt}). The sunspot-group tilt angles were derived using two distinct methods in the DPD catalogue. The first method calculates tilt angles by weighting the spots with their corrected whole spot area, while the second method relies solely on umbral position and area data. Following \cite{Jiao2021}, we adopt the tilt angles obtained from the first approach. As a result, the calculation method for the tilt angle in the two datasets we use is consistent to some extent. Unlike the WJL dataset which uses synoptic magnetograms, sunspots in DPD are often recorded multiple times. To obtain a record of a sunspot group as it crosses the central meridian, we first exclude sunspots with a CMD larger than $10^{\circ}$. If a sunspot group still has multiple records, we retain the one with the smallest CMD. This process results in data for 3408 sunspot groups, with 2119 groups from cycle 23 and 1289 groups from cycle 24.

The white-light-based DPD dataset with the magnetogram-based WJL dataset are the two original datasets that we use for mutual validation.

\subsection{Obtaining the mutually validated datasets}\label{subsec:cscdata}
Discrepancies in tilt angle measurements of ARs/spots across datasets stem from two primary sources. The first is omission bias, which occurs when an AR is included in one dataset but absent in another. For example, decayed ARs without sunspots might be included in a magnetogram-based dataset but excluded in a white-light-based dataset. Also, non-magnetic dark dots might be recognized as sunspots in white-light images but not identified in magnetograms \citep{Tlatov2022}. The second source is measurement inconsistency. Even when both datasets identify the same AR, methodological differences, such as the different grouping of spots/ARs \citep{Baranyi2015,Wang2015} or the effects of magnetic tongues \citep{Poisson2020} and complex structures, like $\delta$-type spots, can produce divergent tilt angle values. To reduce the effect of these measurement errors, we introduce the mutual validation method.

In Step 1, we set an area threshold of 8 MSH to exclude small-area sunspot groups, as they are more likely to be unipolar groups \citep{Baranyi2015} and non-magnetic dots \citep{Tlatov2022}.

\begin{figure}[ht!]
	\centering
	\includegraphics[width=1.0\linewidth]{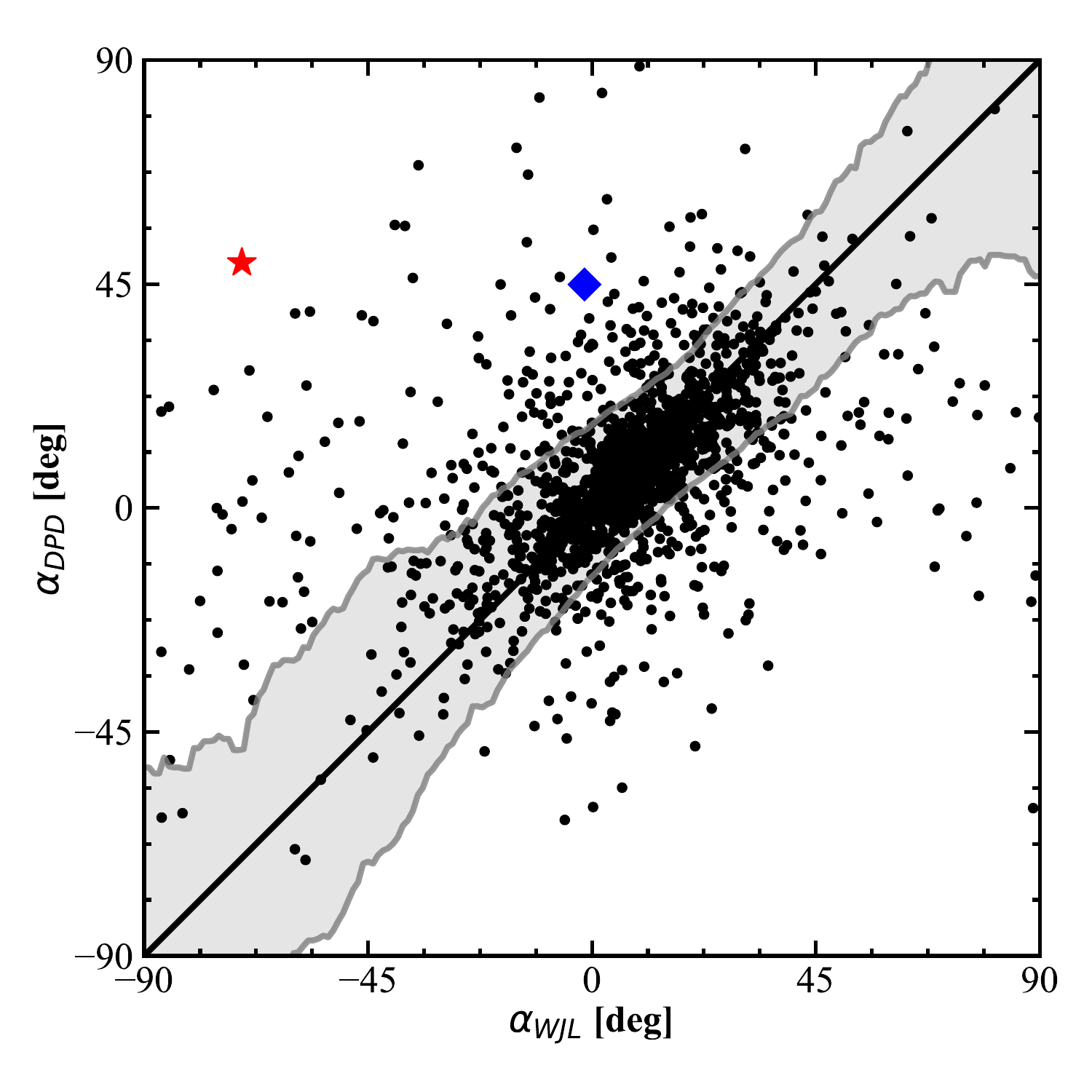}
	\caption{Different tilt values after the matching procedure. The $x$-axis ($y$-axis) represents the tilt angle from the WJL (DPD) dataset. The solid black line is a $y = x$ reference line, and the shaded grey region indicates one standard deviation. The red star and blue diamond symbols represent ARs 9501 and 12673, respectively.}
	\label{fig:1}
\end{figure}

In Step 2, we perform a matching procedure between datasets. We notice that a large portion of records in one dataset do not have counterparts in another dataset. This indicates these records are not reliable and need to be excluded. To do this, we compare the date and heliographic positions of records from both datasets. If two records from the same date have a latitudinal difference smaller than $5^{\circ}$ and a longitudinal difference smaller than $10^{\circ}$, we consider them to be the same AR. In addition to records without counterparts, we also find that some ARs have multiple counterparts due to distant sunspots being assigned to separate groups. There are 156 such ARs, each typically corresponding to 2 or 3 sunspot groups. In total, 439 sunspot groups are associated with these ARs. Since the polarities in such large ARs are often complicated, it is difficult to determine the ``correct" counterpart, so these records are also excluded along with those without counterparts. After step 2, we obtain 1539 paired tilt data. The results are illustrated in Figure \ref{fig:1}.

In Step 3, we exclude data with significantly different tilt values in different datasets. Figure \ref{fig:1} shows the tilt values of the same ARs/sunspot in both datasets after the matching procedure. A given AR/sunspot at a specific time should have comparable tilt angles across different measurements. However, random measurement errors may cause the values to have significant discrepancies. As Figure \ref{fig:1} shows, while most points are near the $y=x$ line, many deviate noticeably. For example, NOAA AR 9501 (marked with a red star in Figure \ref{fig:1}), has a negative tilt angle of $-70^{\circ}.36$ in the WJL dataset and a positive value $49^{\circ}.42$ in the DPD dataset. The significant discrepancy in the tilt values of AR 9501 is primarily due to some plage area surrounding the sunspot, which are only detectable in magnetograms. Another example is AR 12673 (marked with a blue diamond), which has a tilt angle of $44^{\circ}.94$ in the DPD dataset and $-1^{\circ}.5$ in the WJL dataset. This AR is a typical complex sunspot group known for producing continuous flares \citep{Yang2017}. Its $\delta$-type structure results in unreliable measurements from both datasets, as noted by \cite{Jiang2019}. These discrepancies highlight the necessity of mutual validation, as such measurement errors are difficult to detect in a single dataset. The detailed reasons for the discrepancies of other data points will be investigated in the future. 

To exclude data with significant measurement errors in at least one dataset, we calculate the standard deviation of $\alpha_{DPD}$ within the range of $\alpha_{WJL} \pm 10^{\circ}$, where $\alpha_{WJL}$ varies from $-90^{\circ}$ to $90^{\circ}$ in increments of $1^{\circ}$. The shaded grey region in Figure \ref{fig:1} indicates one standard deviation. Points outside this range are excluded. We note that some of the excluded outlier pairs may be reliable in one dataset but not in the other. As a result, this approach could potentially reduce the sample size of reliable data.

Finally, we obtain 1148 paired data, with 624 pairs from cycle 23 and 524 pairs from cycle 24. The mutually validated datasets is available on GitHub\footnote{https://github.com/LangQin01/mutually-validated-tilt-angle-dataset/tree/v1.0.0.} under an MIT License and version 2.0 is archived in Zenodo \citep{dataset}. In the following section, we will investigate three properties of tilt angles using the mutually validated datasets and compare the results with those from the original datasets.

\section{Results} \label{sec:result} 

\subsection{No Significant difference between magnetogram-based and white-light-based tilt angles}
\label{subsec:tiltDifference}

\begin{figure}[ht!]
	\centering
	\includegraphics[width=1.0\linewidth]{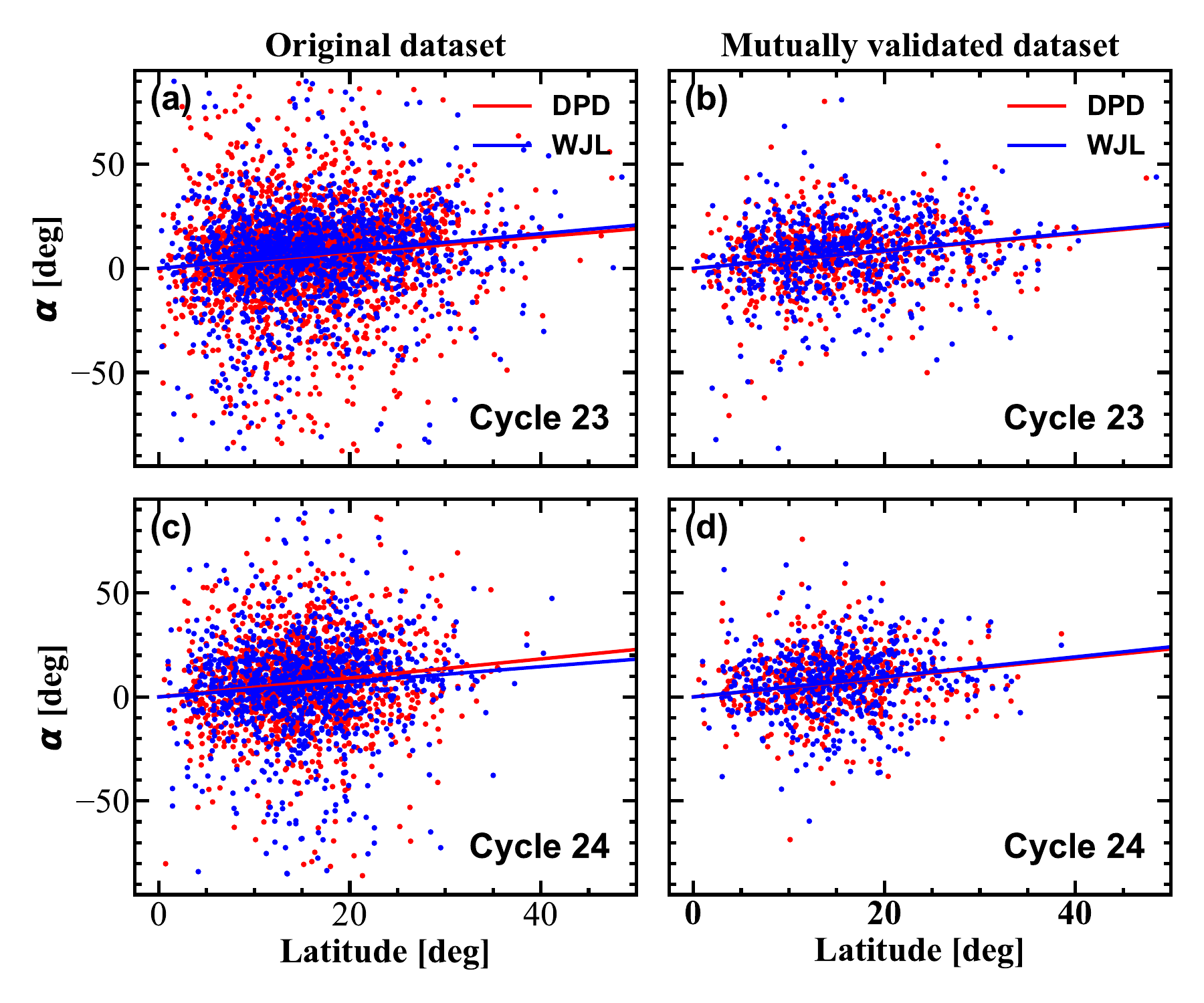}
	\caption{Linear fit to the tilt-latitude relation without binning the data. The left (right) column is based on the original (mutually validated) database. The upper (bottom) panels represent cycle 23 (24). The red (blue) dots are the DPD (WJL) data points, and the red (blue) solid line is the linear fit for the DPD (WJL) data.
		\label{fig:2}}
\end{figure}

To investigate the difference between white-light-based and magnetogram-based tilt angle measurements, we compare two parameters: the average tilt angle $\bar\alpha$ and the tilt coefficient $m$. According to Joy's law, tilt angle increases with increasing latitude \citep{Hale1919}. Different functions describing Joy's law and methods to derive the tilt coefficients have been adopted by previous studies. As proposed by \cite{Jiao2021}, a linear fit to the tilt-latitude relation without binning the data can minimize the effect of the tilt scatter on the uncertainty of the tilt coefficient. Therefore, we perform a direct linear fit $\alpha = m |\lambda|$ without binning the data, as shown in Figure \ref{fig:2}, to derive the tilt coefficient $m$ for different data. The tilt coefficient $m$ and the average tilt angle $\bar\alpha$ based on the original datasets and the mutually validated datasets are presented in Table \ref{tab:1}, respectively.

\begin{deluxetable*}{lCCCC}[t!]
	\tablenum{1}
	\tablecaption{Tilt Angle Parameters\label{tab:1}}
	\tablewidth{0pt}
	\tablehead{
		\colhead{Dataset} & \colhead{Parameter} & \colhead{Cycle 23} & \colhead{Cycle 24} & \colhead{Cycles 23 \& 24}
	}
	\startdata
	Original DPD & \sigma_{\alpha} & 24^{\circ}.32 & 22^{\circ}.04 & 23^{\circ}.48 \\
	{ }& \bar\alpha \pm \sigma_{\bar\alpha} & 6^{\circ}.29 \pm 0^{\circ}.53 & 7^{\circ}.05 \pm 0^{\circ}.61 & 6^{\circ}.58 \pm 0^{\circ}.40 \\
	{ } & m \pm \sigma_{m} & 0.38 \pm 0.03 & 0.46 \pm 0.04 & 0.41 \pm 0.02 \\
	Original WJL & \sigma_{\alpha} & 23^{\circ}.83 & 23^{\circ}.70 & 23^{\circ}.78 \\
	{ }& \bar\alpha \pm \sigma_{\bar\alpha} & 6^{\circ}.48 \pm 0^{\circ}.63 & 5^{\circ}.69 \pm 0^{\circ}.72 & 6^{\circ}.14 \pm 0^{\circ}.47 \\
	{ } & m \pm \sigma_{m} & 0.42 \pm 0.04 & 0.36 \pm 0.04 & 0.40 \pm 0.03 \\
	Mutually validated DPD & \sigma_{\alpha} & 16^{\circ}.18 & 15^{\circ}.23 & 15^{\circ}.76 \\
	{ }& \bar\alpha \pm \sigma_{\bar\alpha} & 6^{\circ}.63 \pm 0^{\circ}.65 & 7^{\circ}.41 \pm 0^{\circ}.67 & 6^{\circ}.99 \pm 
	0^{\circ}.47 \\
	{ } & m \pm \sigma_{m} & 0.42 \pm 0.04 & 0.46 \pm 0.04 & 0.43 \pm 0.03 \\
	Mutually validated WJL & \sigma_{\alpha} & 17^{\circ}.53 & 16^{\circ}.04 & 16^{\circ}.87 \\
	{ }& \bar\alpha \pm \sigma_{\bar\alpha} & 6^{\circ}.81 \pm 0^{\circ}.70 & 7^{\circ}.59 \pm 0^{\circ}.70 & 7^{\circ}.17 \pm 
	0^{\circ}.50 \\
	{ } & m \pm \sigma_{m} & 0.43 \pm 0.04 & 0.48 \pm 0.04 & 0.45 \pm 0.03 \\
	\enddata
	\tablecomments{$\sigma_{\alpha}$: standard deviation of tilt angle; $\bar\alpha$ and $\sigma_{\bar\alpha}$: average tilt angle and the corresponding standard error, respectively; $m$ and $\sigma_{m}$: the slope of the linear fit between tilt angle and latitude and the corresponding fitting uncertainty, respectively.}
\end{deluxetable*}

Comparing the parameters of the original DPD and original WJL datasets in Table \ref{tab:1}, we observe that in cycle 23, the DPD dataset shows $\bar\alpha=6^\circ29 \pm 0^\circ.53$ and $m=0.38 \pm 0.03$, which are smaller than the $\bar\alpha=6^\circ48 \pm 0^\circ.63$ and $m=0.42 \pm 0.04$ of the WJL dataset. However, in cycle 24 and cycles 23 \& 24 combined, DPD dataset shows larger parameters than that of the WJL dataset. As discussed in Section \ref{subsec:cscdata}, both datasets contain a considerable number of records that do not have corresponding records in another dataset and a considerable number of records with differing measurements for the same AR/sunspot at a specific time. We remove these measurement errors through mutual validation. In contrast, the mutually validated datasets consistently indicates that, in all cases, the DPD dataset shows slightly smaller $\bar\alpha$ and $m$ values compared to the WJL dataset. On average, the validated DPD datasets shows $\bar\alpha$ and $m$ values that are only $0^\circ.18$ and $0.02$ smaller, respectively, both of which fall within the standard errors of the means. We may observe another relevant property that all $\bar\alpha$ ($\sim$$7^{\circ}.0\pm0^{\circ}.7$) and $m$ ($\sim$0.43$\pm0.04$) values are larger than those from the original datasets, and much larger than previous view based on white-light observations as presented in Introduction. This suggests that an upward revision of $\bar\alpha$ and $m$ are required for modeling the surface flux transport process \citep{Wang1989b,Jiang2014, Yeates2023}.

The results from the original datasets indicate that white-light tilt values are not necessarily smaller than those from magnetic measurements. Tilt angle datasets from the MWO and KK observations have been widely used in previous studies. Without the reference of magnetograms, tilt angles of sunspots are more likely referred to as sunspots of the same polarity, thus leading to a smaller tilt angle \citep{Wang2015}. These measurement errors significantly contribute to the previous view of smaller white-light-based tilt angles. For the original DPD data herein, contemporaneous magnetograms were referenced to some extent \citep{Baranyi2015}, leading to smaller measurement errors. When measurement errors are further reduced by mutual validation, the difference of tilt angle values between a white-light-based dataset and a magnetogram-based dataset is negligible. The result actually complements results presented in Table 1 of \cite{Wang2015}. For contemporaneous observations in cycle 21 by MW white-light images and magnetograms, without the constraint of the angular separation, both the magnetogram-based average tilt angle $\bar\alpha$ and the tilt coefficient $m$ are significantly larger than those derived from white-light observations. This is consistent with previous view. However, when the angular separation constraint is applied, which closely resembles the first step of mutual validation, the results change remarkably. The discrepancy between the two types of datasets notably decrease, and $\bar\alpha$ and $m$ values increase. The mean value from white-light observations is even larger than that from magnetograms.
	
Additionally, we note that due to the relatively small sample sizes, the uncertainty ranges for $\bar\alpha$ ($\sigma_{\bar\alpha}$) and $m$ ($\sigma_m$) are relatively large. To rigorously verify the relationship between white-light-based and magnetogram-based tilt values, more datasets and datasets with larger sample sizes should be used for mutual validation. Furthermore, we argue that since tilt angle is possibly cycle-dependent \citep{Dasi2010,Jiao2021}, it is not appropriate to compare the tilt angle when the datasets do not overlap in time.
 
\cite{Howard1996b, Wang2015} suggested that for magnetic measurements, the contribution of plage areas in an AR could lead to a larger tilt angle. \cite{Wang2015} presented 4 specific cases, where magnetic measurements were significantly larger than white-light measurements. In the first and third cases, the leading and following sections were assigned to the same polarity, indicating measurement errors that do not reflect the intrinsic tilt values. Such erroneous tilts are expected to be excluded by the mutual validation method. For the second and fourth cases, however, it remains unclear whether the plage areas exhibit a preferred positional relationship relative to sunspots, meaning their contribution may not necessarily increase the tilt angle, like AR 9501 discussed in Section \ref{subsec:cscdata}.

In summary, the discrepancy between white-light-based and magnetogram-based tilt angles is less significant than previously views. Given that cycle 23 has an average amplitude, its $m$ value of $\sim$0.43$\pm0.04$ and $\bar\alpha$ value of $\sim$$7^{\circ}.0\pm0^{\circ}.7$ are proposed as the typical values describing Joy's law and average tilt angles, respectively. We note that the standard errors of these values are still rather large due to the limited sample size.

\subsection{Significant decrease in tilt scatter and its intrinsic dependence on ARs' flux and area}
\label{subsec:tiltScatter}
Tilt scatter is an indicator of the convective buffeting on flux emergence \citep{Weber2013, Schunker2019}. In Table \ref{tab:1} the tilt scatters before and after mutual validation are also included. Table \ref{tab:1} shows that the standard deviations of the tilt angle, $\sigma_\alpha$, are about 23$^\circ$ based on the original datasets. The noisier tilt angle data from KK and MWO have $\sigma_\alpha$ about 30$^\circ$ \citep{Ivanov2012, Jiao2021}. After mutual validation, $\sigma_\alpha$ decreases to approximately 16$^\circ$.

\begin{figure*}[htp!]
	\centering
	\includegraphics[width=1.0\linewidth]{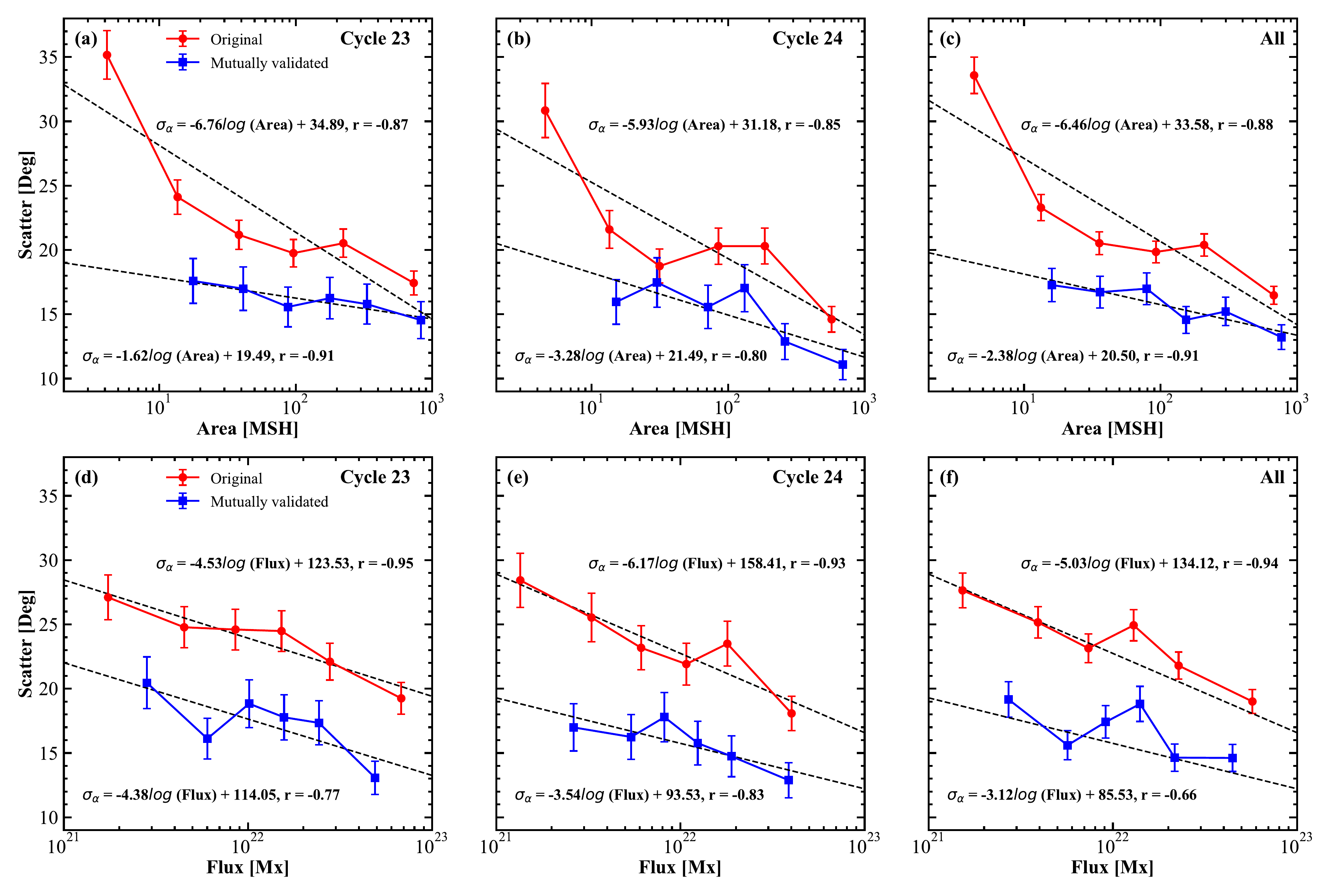}
	\caption{Sunspot area (upper panels) and AR flux (bottom panels) dependence of tilt scatter quantified by standard deviation $\sigma$. The left, middle, and right panels show results for solar cycles 23, 24, and the combined data from cycles 23 and 24, respectively. The red and blue curves are from the original and mutually validated datasets, respectively. The dashed lines are the corresponding fits. The corresponding fitting functions are displayed above the red curves for the original datasets and below the blue curves for the mutually validated datasets. The $r$-values represent the correlation coefficients between $\sigma$ and sunspot area or AR flux.
		\label{fig:scatter}}
\end{figure*}

It has been theoretically \citep{Weber2013} and observationally \citep{Howard1996b, Jiang2014} established that tilt scatter increases with decreasing sunspot area or AR flux. To verify the effectiveness of our mutual validation method in reducing measurement error, we divide the original and mutually validated DPD and WJL datasets into six bins based on logarithmic sunspot area and magnetic flux, respectively. Each bin contains an equal number of data points. We then calculate $\sigma_\alpha$ within each bin and analyze its dependence on sunspot area and AR flux. The results are presented in Figure \ref{fig:scatter}, along with curve fits illustrating the relationship between $\sigma_\alpha$ and area (flux). The x-axis value of each point represents the average value within each bin.

The red curves in Figure \ref{fig:scatter} represent the dependence of $\sigma_\alpha$ on sunspot area (upper panels) and AR flux (lower panels) for the original datasets. We observe that $\sigma_\alpha$ decreases with both sunspot area and AR flux for cycle 23 (left panels), cycle 24 (middle panels), and the combined cycles 23 and 24 (right panels), consistent with previous reports \citep{Fisher1995, Jiang2014}. Furthermore, the $\sigma_\alpha$ values for the mutually validated datasets (blue curves) exhibit significant decreases from the original datasets for each bin. We consider the $\sigma_\alpha$ values for cycle 23 as the typical values since cycle 23 has an average amplitude. The fitted dependence of $\sigma_\alpha$ on area ($A$) is as follows,
\begin{equation}
	\label{eq:tiltScatter_area}	
	\sigma_\alpha=\left\{
	\begin{array}{l l}
		-6.76\log (A)+34.89 & \textrm{Original DPD} \\
		-1.62\log (A)+19.49 & \textrm{Mutually validated DPD}.
	\end{array}
	\right.
\end{equation}
The fitted flux ($F$) dependence of $\sigma_\alpha$ is
\begin{equation}
	\label{eq:tiltScatter_flux}	
	\sigma_\alpha=\left\{
	\begin{array}{l l}
		-4.53\log (F)+123.53 & \textrm{Original WJL} \\
		-4.38\log (F)+114.05 & \textrm{Mutually validated WJL}.
	\end{array}
	\right.
\end{equation}

Based on the mutually validated datasets, the AR flux ($F$) and counterpart of sunspot area ($A$) obeys the relationship of 
\begin{equation}
	F = 10^{20.88} A^{0.57} .
	\label{eq:Flux_Area}
\end{equation}
Sunspot area mainly ranges from 10 MSH to 1500 MSH, with corresponding AR flux ranging from about 3$\times10^{21}$ Mx to about 6$\times10^{22}$ Mx. Based on Equation (\ref{eq:tiltScatter_area}), the original DPD dataset gives $\sigma_\alpha$ values ranging 13$^\circ.42$ to 28$^\circ$.13, while the mutually validated DPD dataset ranges from 14$^\circ.34$ to 17$^\circ$.87. For the original WJL dataset, $\sigma_\alpha$ values range from 20$^\circ$.34 to 26$^\circ$.24, and for the mutually validated WJL dataset, the range is from 14$^\circ$.28 to 19$^\circ$.98. The mutual validation method results in a more significant decrease in scatter for smaller spot size or weaker AR flux by reducing measurement errors. We propose that the lower lines of Equations (\ref{eq:tiltScatter_area}) and (\ref{eq:tiltScatter_flux}) representing the dependence of tilt scatter on sunspot area and AR flux, respectively, originated from solar mechanisms.

Figure \ref{fig:scatter} also shows that the slopes describing the area/flux dependence of $\sigma_\alpha$ have a significant decrease after mutual validation, as presented by Eqs. (\ref{eq:tiltScatter_area}) and (\ref{eq:tiltScatter_flux}). This suggests that the intrinsic $\sigma_\alpha$ may have a weaker dependence on sunspot area or AR flux than previously expected. Although the dependence on sunspot area is weaker, $\sigma_\alpha$ values have high correlation coefficients with area ($|r| \geq 0.8$ and $p \leq 0.05$). However, the correlation coefficients between $\sigma_\alpha$ values and AR flux are weaker, ranging from $r = -0.66$ to $r = -0.83$. The $\sigma_\alpha$ value has a peak around $10^{22}$ Mx, though with some uncertainty. It is unclear if this peak is of solar origin and whether it could serve as a constraint on the debated mechanisms for tilt angles.

\subsection{Dependence of tilt angles on Cycle amplitude and magnetic properties} 
\label{subsec:tiltQuenching}
\subsubsection{Cycle amplitude dependence of tilt angle} \label{subsec:cycle}
Tilt angle being anti-correlated with cycle amplitude (TQ1) works as a nonlinear mechanism to regulate cycle variability \citep{Jiang2020, Talafha2022, Karak2023}. Cycle 24 is about 35\% weaker than cycle 23 based on the 13-month smoothed monthly total sunspot number, making the comparison of tilt angle in cycles 23 and 24 crucial for providing evidence for or against the existence of TQ1.

It has also been found that ARs in stronger cycles emerge at higher latitudes \citep{Solanki2008, Jiang2011}, resulting in larger tilt angles due to Joy's law. Therefore, we compare both $\bar\alpha$ and $m$ (to exclude the latitudinal dependence of tilt angles) in cycles 23 and 24. The results are presented in Table \ref{tab:1}. Cycle 24 is expected to exhibit a larger average tilt angle if TQ1 exists. However, as shown in Table \ref{tab:1}, the original datasets present conflicting results. The original DPD dataset shows significantly larger $\bar\alpha$ and $m$ for cycle 24 compared to cycle 23, while the original WJL dataset shows the opposite trend. This discrepancy highlights the tilt measurement errors in different datasets, rendering neither result reliable and explaining the divergent reported outcomes.

In contrast, the mutually validated datasets provide consistent results. As shown in Table \ref{tab:1}, both the DPD and WJL mutually validated datasets present larger $\bar\alpha$ and $m$ in cycle 24 compared to cycle 23 with statistical significance, that is $\Delta\bar\alpha\geqslant\sigma_\alpha$ and $\Delta m \geqslant\sigma_m$. Based on the empirical relationship between the tilt coefficient $m$ and cycle strength $S_n$ given by \cite{Jiao2021}, $m=-0.00107*S_n+0.61$, $m$-values are 0.42 and 0.48 for cycle 23 ($S_n$=181) and cycle 24 ($S_n$=115), respectively. The values are in good consistency with the mutually validated results. 

The different results given by original and mutually validated datasets demonstrate that the mutual validation method can efficiently reduce the random errors in different datasets and provides evidence supporting the existence of TQ1. 

\begin{figure*}[ht!]
\centering
\includegraphics[width=1.0\linewidth]{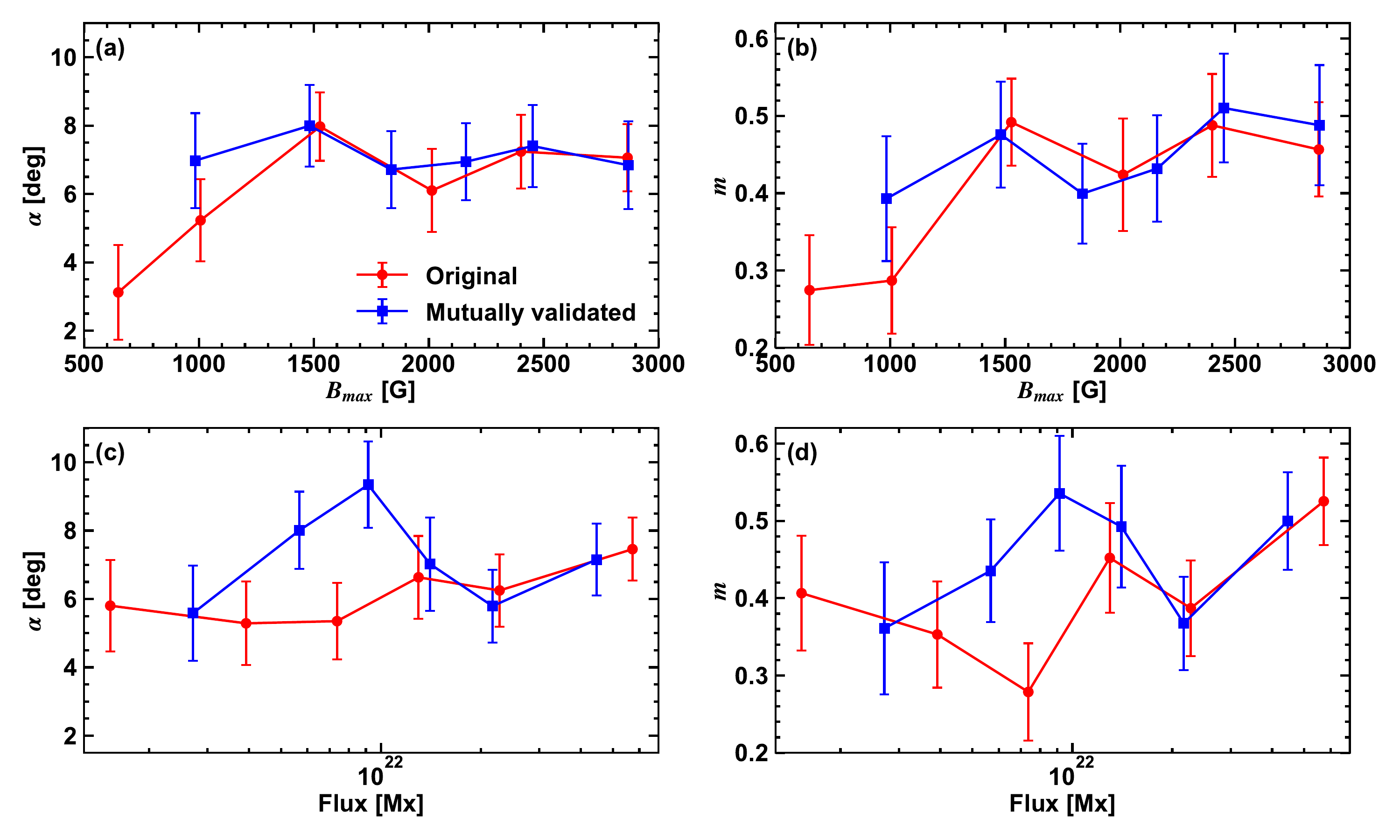}
\caption{Tilt angles and the corresponding flux and $B_{max}$ of ARs. Red dots (blue squares) represent the original (mutually validated) WJL dataset. The top (bottom) panels show the average tilt angle $\alpha$ and the corresponding maximum field strength $B_{max}$ (magnetic flux $\Phi$). The error bars represent one standard error. 
\label{fig:3}}
\end{figure*}

\subsubsection{Magnetic properties dependence of tilt angle} \label{subsec:mag}
Continuous magnetic observations by MDI/SOHO and HMI/SDO allow us to verify the existence and identify the specific dependent parameter of TQ2 (non-monotonic dependence of tilt angle on magnetic properties of ARs). In this subsection, we investigate the relationship between tilt angle and two magnetic properties: the maximum flux density $B_{max}$ and the flux of ARs. We only use the WJL dataset since the DPD dataset does not provide magnetic information. We calculate the average tilt angle $\bar\alpha$ and the linear fit slope $m$ in 6 bins with an equal number of data points (425 for the original dataset and 191 for the mutually validated dataset). The results are illustrated in Figure \ref{fig:3}.

The upper panels in Figure \ref{fig:3} illustrate how $\bar\alpha$ and $m$ vary with $B_{max}$. The red dots, representing the original WJL dataset, show an increase in both parameters up to approximately 1500 Gauss, beyond which they plateau. This suggests a saturation of tilt angle with increasing $B_{max}$ in the original dataset. The rising part looks similar to the corresponding results shown by \cite{Jha2020, Sreedevi2024} but they observed a declining trend in the range from $\sim 2000$ Gauss to $\sim 3000$ Gauss. In contrast, the mutually validated dataset, represented by the blue curves, does not exhibit the same trend. Both $\bar\alpha$ and $m$ show neither saturation nor a clear decrease with increasing $B_{max}$ from $\sim 1000$ Gauss to $\sim 3000$ Gauss. This indicates that the tilt angle has no dependence on the maximum flux density $B_{max}$ of ARs. Additionally, a comparison between the original and mutually validated datasets indicates that ARs having $B_{max}$ smaller than 1 kG are less likely to have sunspots, which is consistent with \cite{Jha2020}.

\cite{Jha2020, Sreedevi2024} connected the theory of the thin flux-tube model for the AR formation to explain their $B_{max}$ dependence of the tilt. Based on the thin flux-tube model, \cite{Fan1994, Fisher1995} provided the tilt angle scaling law $\alpha\propto B_{0}^{-5/4}F^{1/4}$, where $B_{0}$ represents the toroidal field strength at the bottom of the convection zone. However, $B_{0}\propto B_{max}$ is a doubtful assumption. The $B_{0}$ dependence of $\alpha$ cannot be verified by observations. Additionally, the average $B_{max}$ in cycle 23 is 1971.2 Gauss, smaller than 2096.5 Gauss in cycle 24. If the TQ2 mechanism exists, the weaker $B_{max}$ of cycle 23 would be expected to result in a higher $\bar\alpha$, which contradicts the TQ1 result.

The bottom panels in Figure \ref{fig:3} present the same analysis as the upper panels, but with the $x$-axis representing flux. The lower limit of flux, $\sim 3\times10^{21}$ Mx, is in line with previous findings of critical value, beyond which sunspots can be formed \citep{Harvey1993, Hofer2024}. The original and mutually validated datasets show different dependencies of $\alpha$ and $m$ on flux. The original datasets indicate no correlation between tilt angle and flux, consistent with findings from previous studies \citep{Wang1989, Stenflo2012, Jha2020}. For the mutually validated dataset, both $\alpha$ and $m$ initially increase with rising flux, then decrease when the flux exceeds approximately $10^{22}$ Mx, and finally increase for ARs with the largest flux. The separate analyses of cycles 23 and 24 indicate that cycle 23 contributes most to this trend shown in the lower panel of Figure \ref{fig:3}, while cycle 24 shows a weaker and more fluctuating flux dependence. Given the large error bars, we suggest that the flux dependence of the tilt angle remains uncertain based on the data from these two cycles. 

Figure 6 of \cite{Illarionov2015} also shows that for flux in the range from $2 \times 10^{21}$ Mx to $6 \times 10^{21}$ Mx, the tilt has an increasing trend with increasing flux based on the MDI magnetograms. Although their plot indicates uncertainty for higher flux values, they concluded that tilt increases with AR flux/area. Analyzing 517 ARs in cycles 22 and 23 observed by Solar Magnetic Field Telescope at Huairou Solar Observatory, \cite{Tian2003} found that tilt angle increases (decreases) with flux when the flux is smaller (larger) than a critical value of $5 \times 10^{21}$ Mx. Considering the possible underestimation of the strength of strong magnetic fields \citep{Wang2009, Bai2014}, the critical flux is roughly consistent with our results. Given that only a small sample size was analyzed, the decline trend for high AR flux may not be reliable. More data are required to conclusively determine the flux dependence of the tilt angle.

\section{Discussion} \label{sec:discussion}
The upward revision of $\bar\alpha$ (average tilt angle) and $m$ (linear dependence of tilt angle on latitude) have a significant impact on the evolution of the polar field and open flux. The AR tilt angle represents the initial contribution to the axial dipole field, but it is the subsequent flux transport over solar surface that dominates the final contribution to the axial dipole field, corresponding to the polar field at cycle minimum \citep{Wang1991, Jiang2014, Petrovay2020, Wang2021}. The emergence of ARs at lower latitudes results in a larger final contribution to the axial dipole field. ARs that emerge over a certain latitude threshold, determined by the transport parameters \citep{Petrovay2020, Wang2021}, do not contribute to the final axial dipole field. Therefore, the upward revision of $\bar\alpha$ primarily affects ARs at low latitudes. Previous surface flux transport simulations of multiple solar cycles, such as those by \cite{Cameron2010, Jiang2011b, Bhowmik2018}, adopted a low $\bar\alpha$ and a square root dependence of the tilt angle on latitude. This approach led to larger tilt angles for ARs at lower latitudes compared to a linear dependence. Thus, the seemingly overestimated tilt angles due to a square root latitudinal dependence in these studies might be closer to realistic values than the linear case. The larger $\bar\alpha$ and $m$ could be more appropriate, warranting further investigation. 

The downward revision of $\sigma_\alpha$ (the tilt scatter) affects our understanding of solar cycle variability. Previous studies, such as those by \citet{Jiang2014} and \citet{Jiang2020}, used a larger $\sigma_\alpha$ for surface flux transport simulations to investigate the impact of tilt scatter on solar cycle variability. The updated values are now roughly consistent with \citet{Kitchatinov2018}, who also modeled solar cycle variability. Given these updated values, it is crucial to revisit the effects of low tilt scatter to fully understand their implications. Additionally, the traditional measurements of the tilt angle have the approximation of the bipolar magnetic region (BMR), which assumes that both polarities are of equal and regular size. Recently, some studies indicate that the contribution of individual ARs to the final dipole field can differ significantly from this BMR approximation  \citep{Jiang2019, Wang2021, Yeates2020, Wang2024}. Despite this complexity, accurate tilt angle measurements remain important for understanding the stochastic mechanisms that drive solar cycle variability.

Over the past decades, the paradigm for understanding the origin of AR tilt angles and flux emergence has centered around the concept of thin flux tubes embedded in turbulent convection, which rise buoyantly from the base of the convection zone \citep{Fan2021}. Although the revised properties of tilt angles including their average value, scatter, and weak flux dependence, fall within the range of free parameters, these properties alone do not definitively support or refute the thin flux tube paradigm. This paradigm posits that tilt angles are generated below the solar surface. Determining whether tilt angle establishment is a pre-emergence or post-emergence process \citep{Kosovichev2008, Schunker2020} is crucial to validate or challenge this paradigm. Results based on mutually validated tilt angles and derived from daily magnetograms could shed light on the ongoing exploration and potential shift away from the thin flux tube approximation. \citep{Weber2023}.

\section{Conclusions} \label{sec:conclusion}
To address the controversial results in previous studies regarding sunspot tilt angles, we have developed a novel method for mutually validating different datasets to reduce random measurement errors in this paper. This approach effectively enhances the reliability of the tilt angle data by identifying the common data where both datasets provide comparable tilt angle values for the same sunspot group/AR. Consequently, the results based on the mutually validated data more accurately reflect the intrinsic solar mechanisms influencing tilt angles, rather than artifacts introduced by measurement errors in original datasets. To our knowledge, this could be the first time to apply the method to reduce measurement errors in tilt angle data.

The main properties of tilt angles investigated in the paper are summarized as follows.
\begin{enumerate}
	\item The differences between magnetogram-based and white-light-based tilt angles are not as significant as once thought. Earlier views suggesting a significant difference between white-light-based and magnetogram-based tilt angles might result from measurement errors. Although the tilt angle of each cycle varies with cycle amplitude, the average tilt angle ($\bar\alpha$) and tilt coefficient ($m$) of cycle 23 can be regarded as typical since this cycle has an average amplitude. Their values are $\bar\alpha \approx7^\circ$ and $m \approx 0.43$, which are larger than earlier accepted values applied in surface flux transport models \citep{Wang1989b, Jiang2014, Yeates2023}. 
	\item The standard deviation of tilt angle, $\sigma_\alpha$, is $\sim 16^\circ-18^\circ$, which is much less than earlier reported values. While the early values are contributed by both measurement errors and intrinsic solar mechanisms, our result is mainly contributed by the latter case. The ratio between $\sigma_\alpha$ and $\bar\alpha$ is less than three. The values of $\sigma_\alpha$ have a strong relation with sunspot area $A$ and with AR flux $F$. Their relationships are $-1.62\log (A)+19.49$ and $-4.38\log (F)+114.05$, respectively. 
	\item The mutually validated magnetogram-based and white-light-based tilt angle in cycle 24 is statistically larger than that in cycle 23. Given the weakness of cycle 24, the result supports previously reported tilt quenching, wherein stronger (weaker) cycles have smaller (larger) average tilt angles. The mutually validated dataset indicates that tilt angles do not depend on the maximum magnetic field strength of ARs. And there is a weak non-monotonic relationship between tilt angle and magnetic flux, but accompanied by large error bars.  
\end{enumerate}		

New datasets continuously emerge with advancements in solar observations and the ongoing compilation of historical data. However, random errors are still inevitably included. If these errors are not reduced, we will probably continue to obtain controversial results about tilt angle like before. This study represents a first step in reducing measurement errors in tilt angle datasets. Further work within the community is essential, particularly the mutual validation of additional tilt angle datasets with larger sample sizes, to further improve the accuracy and reliability of the results. Ultimately the intrinsic properties of sunspot tilt angles can be clearly understood.

%

\begin{acknowledgments}

The research is supported by the National Natural Science Foundation of China (grant Nos. 12425305, 12350004, and 12173005) National Key R\&D Program of China (grant No. 2022YFF0503800), and the Climbing Program of NSSC (E4PD3001).

\end{acknowledgments}

\bibliography{sample631}{} 
\bibliographystyle{aasjournal}


\end{document}